\documentclass[usegraphicx,usenatbib,useapjfonts,apj]{emulateapj}
\usepackage{amssymb}

\def\lesssim{{_ <\atop{^\sim}}}
\def\grtsim{{_ >\atop{^\sim}}}

\def\ms{\mbox{$M_{\rm s}$}}
\def\mh{\mbox{$M_{\rm h}$}}
\def\mtran{\mbox{$M_{\rm tran}$}}
\def\dotM{\mbox{$\dot{M}_s/M_s$}}
\def\dotMz{\mbox{$\dot{M}_s(z)/M_s(z)$}}
\def\Mmax{\mbox{$M_{\rm hp}$}}

\def\msun{\mbox{$M_{\odot}$}}
\def\lcdm{\mbox{$\Lambda$CDM}}
\def\gsmf{\mbox{$GSMF$}}
\def\gsmfs{\mbox{$GSMF$s~}}

\def\apj{\mbox{ApJ}}
\def\apjs{\mbox{ApJSS}}
\def\mnras{\mbox{MNRAS}}
\def\apjl{\mbox{ApJL}}

\def\aap{\mbox{A\&A}}

\def\spose#1{\hbox to 0pt{#1\hss}}
\newcommand\lsim{\mathrel{\spose{\lower 3pt\hbox{$\mathchar"218$}}
     \raise 2.0pt\hbox{$\mathchar"13C$}}}
\newcommand\gsim{\mathrel{\spose{\lower 3pt\hbox{$\mathchar"218$}}
     \raise 2.0pt\hbox{$\mathchar"13E$}}}

\slugcomment{Accepted}

\shorttitle{Galaxy downsizing}
\shortauthors{Firmani \& Avila-Reese}

\begin{document}


\title{Galaxy downsizing evidenced by hybrid evolutionary tracks}


\author{C. Firmani\altaffilmark{1}}
\affil{INAF-Osservatorio Astronomico di Brera, via E.Bianchi 46, I-23807
Merate, Italy}
\email{firmani@merate.mi.astro.it,avila@astro.unam.mx}
\and
\author{V. Avila-Reese}
\affil{Instituto de Astronom\'{\i}a, Universidad Nacional Aut\'onoma de M\'exico,
A.P. 70-264, 04510, M\'exico, D.F., M\'exico}

\altaffiltext{1}{Also Instituto de Astronom\'{\i}a, Universidad Nacional Aut\'onoma de M\'exico,
A.P. 70-264, 04510, M\'exico, D.F., M\'exico}


\begin{abstract}

The stellar--dark halo mass relation of galaxies at different redshifts, \ms(\mh,$z$), 
encloses relevant features concerning their physical processes and evolution. 
This sequence of relations, defined in the range $0< z < 4$, together with average $\Lambda$ 
cold dark matter (\lcdm) halo mass aggregation histories (MAHs) are used here for inferring 
{\it average} \ms\ growth histories, the Galaxian Hybrid Evolutionary Tracks (GHETs), 
where 'hybrid' remarks the combination of observational (\ms) and theoretical (\mh) ingredients. 
As a result of our approach, a unified picture of stellar and halo mass buildup, population 
migration, and downsizing of galaxies as a function of mass is presented.

The inferred average \ms\ growth histories (GHETs) of highest and lowest mass galaxies are definitively 
quite different from the average MAHs, \mh($z$), of the corresponding dark 
halos. Depending on how a given \mh($z$) compares with the mass at which the \ms-to-\mh\ ratio 
curve peaks at the epoch 
$z$, \Mmax($z$), two evolutionary phases are evidenced: (i) galaxies in an active regime 
of \ms\ growth when $\mh < \Mmax$, and (ii) galaxies in a quiescent or passive regime 
when $\mh > \Mmax$. The typical \ms\ at which galaxies transit from the active (star-forming) to the 
quiescent regime, \mtran, increases with $z$, log(\mtran/\msun)$\approx 10.30 + 0.55z$,  
making evident a {\it population downsizing} phenomenon.
This result agrees with independent observational determinations based on the evolution of 
the galaxy stellar mass function decomposition into blue and red galaxy populations. 
The specific star formation rate, SSFR, predicted from the derivative of the GHET is 
consistent with direct measures of the SSFR for galaxies at different redshifts, though 
both sets of observational inferences are independent. 
The average GHETs of galaxies smaller than \mtran\ at $z=0$ 
(\ms$\approx 10^{10.3}$ \msun, \mh$\approx 10^{11.8}$ \msun) did not reach the 
quiescent regime, and for them, the lower the mass, the faster the later \ms\ growth rate
({\it downsizing in SSFR}). 
The GHETs allow us to predict the transition rate in the number density
of active to passive population; the predicted values agree with direct estimates of the
growth rate in the number density for the (massive) red population up to $z\sim 1$.
We show that \lcdm--based models of disk galaxy evolution, including 
feedback-driven outflows, are able to reproduce the low-mass side of the \ms--\mh\ 
relation at $z\sim 0$, but at higher $z'$s strongly disagree with the GHETs: models 
do not reproduce the strong downsizing in SSFR and the high SSFR of low mass galaxies. 

\end{abstract}




\section{Introduction}

According to the popular hierarchical clustering $\Lambda$ Cold Dark Matter (\lcdm) scenario, galaxies 
form and grow inside evolving dark matter haloes. A central question emerges then about what is the 
galaxy stellar mass, \ms, associated on average with a given halo of mass \mh, i.e., the \ms--\mh\ relation. 
The change with redshift of this relation, \ms(\mh,$z$), resumes the key astrophysical processes of 
galaxy stellar mass assembly in the context of the \lcdm\ scenario.

With the advent of large galaxy surveys in the last years, a big effort has been made in constraining 
the local \ms--\mh\ relation (1) directly by estimating halo masses with galaxy-galaxy weak lensing, 
with kinematics of satellite galaxies or with X-ray studies; and (2) indirectly by linking the observed statistical 
galaxy properties (e.g., the galaxy stellar mass function \gsmf, the two--point correlation function, galaxy 
group catalogs) to the theoretical Halo Mass Function (hereafter $HMF$; for recent reviews 
on all of these methods see Moster et al. 2010; Behroozi et al. 2010;  More et al. 2010, 
and more references in these papers). 
The latter approach is less direct than the former one,  but it allows to cover larger mass ranges, 
and it can be extended up to the redshifts where observed \gsmfs\ are reported
(for recent results see Conroy \& Wechsler 2009; Moster et al. 2010; Wang \& Jing 2010; 
Behroozi et al. 2010). 

In Moster et al. (2010) and Behroozi et al. (2010; hereafter BCW10), the \ms(\mh,$z$) functionality  
has been constrained up to $z\approx 4$. In each case, different observational data sets 
for the \gsmfs\ and different methods for statistically assigning halo masses to the galaxies 
were used. The \ms--\mh\ relation at $z\sim 0$ is similar in both works (see also Baldry et al.
2008; Drory et al. 2009; Guo et al. 2010; Wang \& Jing 2010). However, 
the change with $z$ of this relation, \ms(\mh,$z$), particularly for $z \grtsim 1$, is 
different in both cases.

Although the uncertainties in the inferences of \ms(\mh,$z$) are still significant, the 
current determinations can be used for preliminary explorations of the galaxy mass assembly process. 
An original approach has been introduced by Conroy \& Wechsler (2009; hereafter CW09), who 
proposed a parametric form for \gsmf\ as function of $z$ constrained by both some observational
reports of this function and the star formation rate (SFR)--\ms\ relations to $z\sim 1$. 
Then, the cumulative \gsmfs\ at each $z$ were matched to the cumulative 
$HMF$s in order to infer the \ms--\mh\ relations at different $z'$s (abundance matching formalism, 
the simplest of the indirect methods; e.g., Marinoni \& Hudson 2002, Kravtsov et al. 2004, 
Vale \& Ostriker 2004; see \S 2). Finally, the obtained \ms--\mh\ relations at different 
$z'$s were connected by using simple parameterizations of the average halo mass aggregation 
histories (MAHs) in order to infer the average stellar mass buildup of galaxies as a function 
of mass. 

The previous approach has the advantage 
(1) of being flexible enough as to allow for explorations covering large mass and redshift ranges, and 
(2) of providing a bridge between observational results and theoretical work. One of the bases of 
this approach is the statistical abundance matching formalism, which has shown to give robust results 
in agreement with direct inferences of the local \ms--\mh\ relation or with other indirect inferences 
based on observations of the two-point correlation functions or galaxy group catalogs 
(see Moster et al. 2010; BCW10; More et al. 2010; Dutton et al. 2010a). 

From the purely empirical point of view, recent studies are posing a new conception in our 
understanding of galaxy stellar mass assembly. For example, Drory \& Alvarez (2008) used 
empirical \gsmfs\ up to $z\sim 5$ (Drory et al. 2005) to infer the \ms\ assembly of large 
galaxies, and in combination with available observations of the SFR--\ms\ relation at different 
$z'$s, they constrained the contribution of star formation (SF) and merging to stellar mass build 
up in galaxies. From this and other studies (e.g., Bundy et al. 2006,2009; Hopkins et al. 2007;
Pozzetti et al. 2010), the trend of downsizing (Cowie et al. 1996) is clearly shown in the sense 
that the mass at which the SFR starts to drop strongly decreases with time (from 
$\ms\sim 10^{12}$ \msun\ at $z>4$ to $\ms\sim 10^{10.9}$ \msun\ at $z\sim 0.5$ according
to Drory \& Alvarez 2008). 
This result combined with observationally inferred merger rates, led the latter authors to 
conclude that the (massive) red sequence is built up from top to down. This phenomenon 
was originally dubbed as {\it 'archaeological downsizing'} (Thomas et al. 2005; see also for 
related results e.g., Daddi et. al 2004,2007; Drory et al. 2004; Bundy et al. 2005,2006; 
Conselice et al. 2007; Marchesini et al. 2009; Perez-Gonzalez et al. 2008 and more 
references therein), and it seems that its main driver are internal galaxy processes that
efficiently quench SF in massive galaxies (e.g., Bundy et al. 2006; Peng et al. 2010). 

The latter result has been confirmed by recent decompositions of the \gsmf\ by galaxy type 
(based on color, SFR, morphology, etc.) up to $z\sim 1$ (e.g., Borch et al. 2006; Bell et al. 2007; 
Drory et al. 2009; Ilbert et al. 2010; Pozzetti et al. 2010; for a compilation of early 
observations, see Hopkins et al. 2007). A systematic result obtained in these works is that  
blue/late-type (active) galaxies migrate to the red/early-type (passive) population involving 
a sequence of masses that decreases with time, a result in line with the archaeological 
downsizing phenomenon. The environment plays also an important role. Recent observational 
studies, where the \gsmf\  has been 
divided not only by galaxy types but also by environment, show that the population migration 
happens more efficiently and earlier in denser environments (e.g., Peng et al. 2010). 

At lower masses the inferences of the \ms\ assembly are difficulted by the 
incompleteness limit of the samples as higher is $z$. However, at least for 
$z\lesssim 1$, current studies show that the specific SFR (SSFR) of low-mass (blue) 
galaxies is surprisingly high, and higher than the SSFRs of more massive galaxies 
(e.g., Bauer et al. 2005; Noeske et al. 2007; Zheng et al. 2007; Bell et al 2007; 
Elbaz et al. 2007; Chen et al. 2009; Damen et al. 2009a,b; Santini et al. 2009; Oliver et al. 2010;
Rodighiero et al. 2010). 
This {\it 'downsizing in SSFR' of low-mass galaxies} seems to imply a delayed \ms\
buildup as lower is the mass, something difficult to explain by the moment in \lcdm-based
models (for models, discussions, and more references see Noeske et al. 2007; 
Fontanot et al. 2009; Firmani, Avila-Reese \& Rodr\'iguez-Puebla 2010).

Based on the \lcdm\ scenario and the new observational inferences
discussed above, a unified picture of stellar and halo mass buildup of galaxies 
of all sizes can be developed.
By using \ms--\mh\ relations constrained by BCW10 from $z\approx 0$
to $4$, we generate an analytical \ms(\mh,$z$) relationship, which is
continuous and differentiable at any $z$. This sequence of
\ms--\mh\ relations in $z$ is combined with average halo MAHs in order to: 

\begin{enumerate}
\item calculate the {\it average} stellar mass buildup tracks of galaxies, \ms($z$), 
from $z=4$ up to now (hereafter, galaxian hybrid evolutionary tracks, GHETs) along with
their halo MAHs, \mh($z$); 
\item calculate the average SSFR histories of galaxies from the corresponding specific stellar mass 
growth rate histories, \dotMz; 
\item determine the typical mass \mtran\ at each $z$ that marks the transition from active to 
passive galaxy population (in terms of \ms\ growth) 
\item infer the transition rate in the number density of active to passive galaxies at each $z$.
\end{enumerate}

Our approach differs from CW09 in several aspects. These authors used
observational inferences of the SSFR vs \ms\ at different $z'$s for constraining
the evolution of their proposed \gsmf\ (up to $z\sim 1$), which is a partial input 
of their approach for calculating \ms(\mh,$z$). Instead, we use more recent and updated 
direct constraints for this relationship (BCW10), which extends up to $z\approx 4$,
and then predict the SSFR histories. In CW09 a simple parameterization for the average 
MAHs was used, while we generate them by using an extended Press-Schechter formalism
that gives results similar to those of large N-body cosmological simulations.
We differ from CW09 also in several aims and results.  

The inferences of the quantities listed above constrain in general the multiple 
astrophysical processes participating in the formation and evolution of galaxies. 
In particular, we will compare the inferred evolutionary tracks (GHETs) to predictions of 
standard $\lcdm$-based semi-numerical models of disk galaxy evolution and explore whether 
the so-called downsizing in SSFR is really an issue for models.
                   
In \S 2, the method to calculate the GHETs is explained.
The inferred average GHETs are presented in different diagrams in \S 3. In \S\S 3.1 
and 3.2 the evolution of the SSFR as a function of mass, the evolution of the transition mass 
\mtran, and the flow of active to passive galaxies at each epoch, are shown and compared with direct 
observational estimates. The reliability of our approach and the results are discussed in \S 3.3. 
In \S 4, \lcdm-based models of disk galaxy evolution are used to calculate evolutionary tracks for 
low-mass galaxies and see whether they are able to reproduce or not the GHETs, in particular the SSFR 
downsizing phenomenon. A summary and our conclusions are given in \S 5.

\section{The Method}

Our inference of the galaxy stellar mass buildup as a function of mass 
(called here GHETs) consists of two main steps: (1) we find an analytical functionality for 
\ms(\mh,$z$), and (2) after generating the halo MAHs, \mh($z$), we combine
them with \ms(\mh,$z$) to infer the GHETs, \ms($z$).  

\begin{figure*} 
\vspace{11.1cm}
\includegraphics{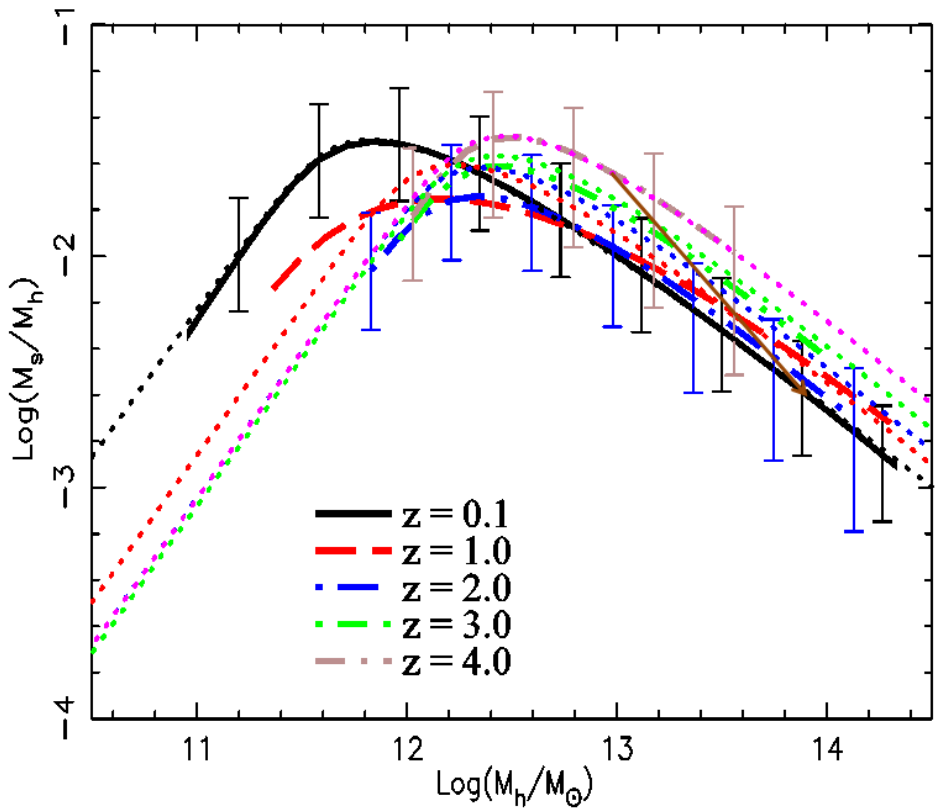}
\caption{\ms/\mh\ vs. \mh\ relations at $z=0.1$, 1, 2, 3, and 4 from BCW10 
with the respective uncertainties. 
The brown arrow shows the evolution of a passive galaxy (constant \ms) from $z=4$,
where its initial conditions coincide with the corresponding isochrone, to $z=0$.
The fact that such arrow reaches the $z=0$ isochrone shows that at these high masses 
galaxies evolve passively (see the text).
The \ms--\mh\ relations given by the \ms(\mh,$z$) relationship proposed here are 
represented by dotted lines with the same color code as the BCW10 curves.
}
\label{galefficiency-b}
\end{figure*}

\subsection{The Stellar Mass--Halo Mass Relations at Different Redshifts}

Our starting point is the determination of \ms(\mh,$z$) up to the highest redshift as possible. 
As mentioned in Section 1, such a relationship has been recently constrained up to $z\sim 4$ 
by using the statistical formalism of matching one-to-one the cumulative theoretical \lcdm\ $HMF$ 
to the observed cumulative \gsmf\ at different $z'$s:
$n_h(>\mh) = n_g(>\ms)$ (e.g., Moster et al. 2010; BCW10).  
Here we will use the results from BCW10, who started from an assumed parametric expression for
\ms(\mh,$z$) and explored the parameter space, including uncertainties and sample variance, 
with a Monte Carlo Markov Chain technique to fit jointly the \gsmfs\ observed at different $z'$s
(they actually carried out this analysis separately for two redshift ranges: $0\lesssim z\lesssim 1$
and $1\lesssim z\lesssim 4$). 
The authors have estimated and carefully  taken into account all kind of uncertainties, which 
introduce significant scatter in the \ms--\mh\ relations and affect even the shape of these relations 
(see  \S\S 3.3 for a discussion). 

The observational input used in BCW10 is the total local \gsmf\ obtained in Li \& White (2009) from Sloan 
Digital Sky Survey (SDSS) data, and the total \gsmfs\ at higher redshifts (up to $\sim 4$) homogeneously 
obtained in P\'erez-Gonz\'alez et al. (2008). For the halo mass function, BCW10 used the results 
from N-body cosmological simulations by Tinker et al. (2008). Since these $HMF$s refer only to 
distinct halos, a correction by the sub-halo abundance was introduced (see details in BCW10). 
The mass of the subhalo is fixed as that one at the time of accretion into the halo, $M_{\rm acc}$, but the 
satellite galaxy's \ms\ is associated with the current \gsmf, i.e., it is implicitly assumed that 
the satellite's \ms\ will continue to evolve in the same way as for centrals of 
halo mass $M_{\rm acc}$ (the effect of fixing instead the satellite mass \ms\ at the redshift 
of accretion has virtually no effect on the overall \ms--\mh\ relation as checked in BCW10).
Thus, the \ms(\mh,$z$) relationship used here refers to the {\it overall} galaxy population
at each epoch, and this {\it includes} satellite galaxies, though under the assumptions mentioned 
above for this sub-population. On the one hand, there are some hints that the \ms--\mh\ relation 
of only satellite galaxies should not differ significantly from the overall relation (Wang et al. 2006).
On the other hand, satellites are not a dominant population at any epoch, in particular at 
high redshifts. Therefore, even if its \ms--\mh\ relation is quite peculiar, its effect over
the average \ms--\mh\ relation is expected to be negligible. 
For more details on the assumptions and role of uncertainties in the inference of the 
\ms(\mh,$z$) relationship we refer the reader to the original work by BCW10 (see also \S\S 3.3).

It is worth mentioning that BCW10 refer to their \ms--\mh\ relations as corresponding
to "central" galaxies. This should be understood in the sense that for each (sub)halo 
is considered only the \ms\ of one galaxy, the central one, but not in the
sense that the satellite galaxy population has been excluded from the analysis (see
above). A (sub)halo may contain more galaxies besides the central one (satellites). 
Therefore, the total stellar mass within the (sub)halo will be larger if we account for 
satellites. In BCW10 it was shown that the inclusion of satellites results in a "total" 
\ms/\mh\ ratio sligthly larger than the "central" one at small masses but significantly 
larger at high (group and cluster) masses (BCW10; see also Yang, Mo \& van den Bosch 2009). 
Some of this "excess" stellar mass is expected 
to end actually in the central galaxy due to satellite infall and/or tidal stripping of 
stars (other fractions of this stellar mass may end in the stellar halo or remain in 
the surviving satellites); this could modify the obtained overall \ms--\mh\ relation\footnote{
According to the analysis of Yang 
et al. (2009), for halos smaller than $\mh\lesssim 10^{13}$ h$^{-1}$\msun\ at $z\sim 0$, 
the mass fraction of stars accreted by the central galaxy is negligible, indicating that 
the latter cannot have grown substantially due to the accretion of satellite galaxy's stars; 
rather their growth is dominated by individual evolution (in situ SF). For halos larger 
than $\mh\approx 10^{13}$ h$^{-1}$\msun, the central galaxy \ms\ may increase lately in a 
significant fraction by satellite accretion (dry mergers); another fraction 
of the accreted stars end in the stellar halo (intracluster stars).}. 
In general, we do not expect that the overall \ms--\mh\ relations inferred by BCW10 will be
significantly affected by the (unknown) physics of satellite galaxies, at least
for $\mh\lesssim 10^{13}$ \msun\ at any epoch, and for any mass at $z>>0$.

Figure \ref{galefficiency-b} shows the \ms-to-\mh\ ratio as a function of \mh\ at $z=$ 0.1, 1.0, 
2.0, 3.0, and 4.0 reported in BCW10 (solid, dashed, dot-dashed, dot-dot-dashed, and dot-dashed-dashed 
lines) with the respective 1$\sigma$ uncertainties (error bars, showed only for $z=0.1, 2,$ and 4). 
Several features regarding galaxy mass assembly can be anticipated from this figure:

1) In the $0 \leq z \leq 4$ range the peak of the curves shifts from log(\Mmax/\msun)$\sim 11.8$ 
to $12.5$, while log(\ms/\mh)$_p$ first declines from $-1.5$ at $z\approx 0$ to $-1.7$ at 
$z\approx 2$, and then recovers the value $-1.5$ at $z\approx 4$. Taking into account the uncertainty, 
the shape of the relations is conserved. Around log(\mh/\msun)$\sim 12.5\pm 0.3$ the curves 
cross each other.

2) On the left side of the diagram, below the crossing masses (in particular from $z=1$ to $z=0.1$),  
for a given \mh, the \ms-to-\mh\ ratio increases as $z$ decreases, which reveals a strong late 
growth of \ms\ for individual galaxies. Hence, this side is expected to be populated mainly 
by {\it actively} growing (blue) galaxies. 

3) Just the opposite happens on the right side, above the crossing masses, 
which means that for individual galaxies, \ms\ brakes its growth while \mh\ continues growing. 
To illustrate this point, we plot in Fig. \ref{galefficiency-b} the evolution (brown arrow) 
of a completely passive galaxy of constant mass $\ms\approx10^{11.35}$  that starts in the 
isochrone $z=4$ with $\mh = 10^{13} \msun$ and evolves until \mh\ grows to $\approx 10^{13.9} \msun$ 
at $z=0$ according to a typical \lcdm\ MAH (see \S\S 2.2) for this mass. If the galaxies of the sample that allow 
us to find the isochrones would be growing in \ms, then the arrow would diverge from the $z=0$ isochrone. 
Therefore, the fact that the arrow reaches the $z=0$ isochrone exactly reveals that this side of the 
diagram is dominated by galaxies with {\it passive} evolution.

The previous considerations confirm the idea that \ms(\mh,$z$) resumes the key astrophysical processes 
of galaxy stellar mass assembly and therefore it contains information about the evolution of 
individual galaxies. Our aim now is to explore the implications of the \ms(\mh,$z$) relations
shown in Fig. \ref{galefficiency-b}. As we are interested in the properties of \ms($z$) up to 
its second derivative (SSFR and the rate of change of SSFR), we cannot use the double parameterization 
in $z$ given by BCW10 in the ranges $0<z<1$ and $1<z<4$, respectively, because discontinuities 
arise around $z=1$. Then, starting from the data of Fig. \ref{galefficiency-b}, we introduce a 
modified parameterization that is continuous in $z$. The following equations, adapted 
from BCW10, summarize the model to be used here for $0<z<4$:
\begin{displaymath}
 \log(\mh(\ms)) = \hspace{0.65\columnwidth}
 \end{displaymath}
 \vspace{-3ex}
\begin{equation}
\quad \log(M_1) + \beta\,\log\left(\frac{\ms}{\ms_{,0}}\right) +
 \frac{\left(\frac{\ms}{\ms_{,0}}\right)^\delta}
{1 + \left(\frac{\ms{,0}}{\ms}\right)^{\gamma}} - \frac{1}{2}.
\label{Mh_Ms}
\end{equation}
The dependence on $z$ is introduced in the parameters of equation 
(\ref{Mh_Ms}) as:
\begin{eqnarray}
\log(M_1(a)) & = & M_{1,0} + M_{1,a} \, (a-1), \nonumber\\
\log(\ms_{,0}(a)) & = & \ms_{,0,0} + \ms_{,0,a} \, (a-1)+ \chi \left( z \right), \nonumber \\
\beta(a) & = & \beta_0 + \beta_a \, (a-1),\\ 
\delta(a) & = & \delta_0 + \delta_a \, (a-1), \nonumber \\
\gamma(a) & = & \gamma_0 + \gamma_a \, (a-1),  \nonumber
\label{zevolution}
\end{eqnarray} 
where $a=1/(1+z)$ is the scale factor.
The function $\chi \left( z \right)$ controls the change with $z$ of the peak value 
(ordinate) of the curves in Fig. \ref{galefficiency-b}. 
If $\chi \left( z \right) = 0$, then the curve peak ordinate is independent of $z$. 
We chose $\chi \left( z \right)$ in order to reproduce roughly the evolution of the peak found 
in BCW10. The first two Eqs. (2) control the position of the \ms/\mh\ peak at each $z$ 
(the value of \Mmax[$z$]), while the last three ones control the shape of the \ms/\mh--\mh\ 
curves. We fix the parameters in Eqs. (1) and (2) expecting (i) to reproduce well the BCW10 
(\ms/\mh)--\mh\ relations at the redshift extremes ($z\approx 0$ and $z\approx 4$, see 
Fig. \ref{galefficiency-b}), 
(ii) to keep the generated relations at all $z'$s within the 1$\sigma$ uncertainty given in BCW10,
and (iii) to have the values of most of the parameters reasonably inside the uncertainties 
of BCW10 (see their Table 2).
The best set of parameters that we have found is reported in Table 1 (for 
comparison, we reproduce there also the best fit values from BCW10 for their 
$0< z< 1$ case with free systematic parameters $\mu$ and $\kappa$). For the function
$\chi$($z$), we use: 
\begin{equation}
\chi \left( z \right) = -0.181 z (1-0.378 z (1-0.085 z)). 
\end{equation}
The agreement of our model with the BCW10 results can be appreciated in 
Fig. \ref{galefficiency-b} (dotted curves with the same color code as
the results from BCW10 at $z=0, 1, 2, 3, 4$).


\begin{table}
\begin{center}
\caption{Best parameters for the \mh(\ms,$z$) functionality}
\label{parameters}
\begin{tabular}{lrr}
\hline
\hline
Parameter & BCW10 & This Work \\
  & $0<z<1$ & $0<z<4$ \\
\hline
$\ms_{,0,0}$  &  $10.72^{+0.22}_{-0.29}$  &  $10.70$ \\
$\ms_{,0,a}$  &    $0.55^{+0.18}_{-0.79}$  &  $-0.80$  \\
$M_{1,0}$        &   $12.35^{+0.07}_{-0.16}$  & $12.35$  \\
$M_{1,a}$        &     $0.28^{+0.19}_{-0.97}$   & $-0.80$  \\
$\beta_0$          &     $0.44^{+0.04}_{-0.06}$   & $0.44$   \\
$\beta_a$          &     $0.18^{+0.08}_{-0.34}$   & $0.00$   \\
$\delta_0$         &     $0.57^{+0.15}_{-0.06}$   & $0.48$   \\
$\delta_a$         &     $0.17^{+0.42}_{-0.41}$   & $-0.15$   \\
$\gamma_0$     &     $1.56^{+0.12}_{-0.38}$   & $1.56$   \\
$\gamma_a$     &     $2.51^{+0.15}_{-1.83}$   & $0.00$   \\
\hline
\end{tabular}
\end{center}
\end{table}


\begin{figure*} 
\vspace{12cm}
\includegraphics{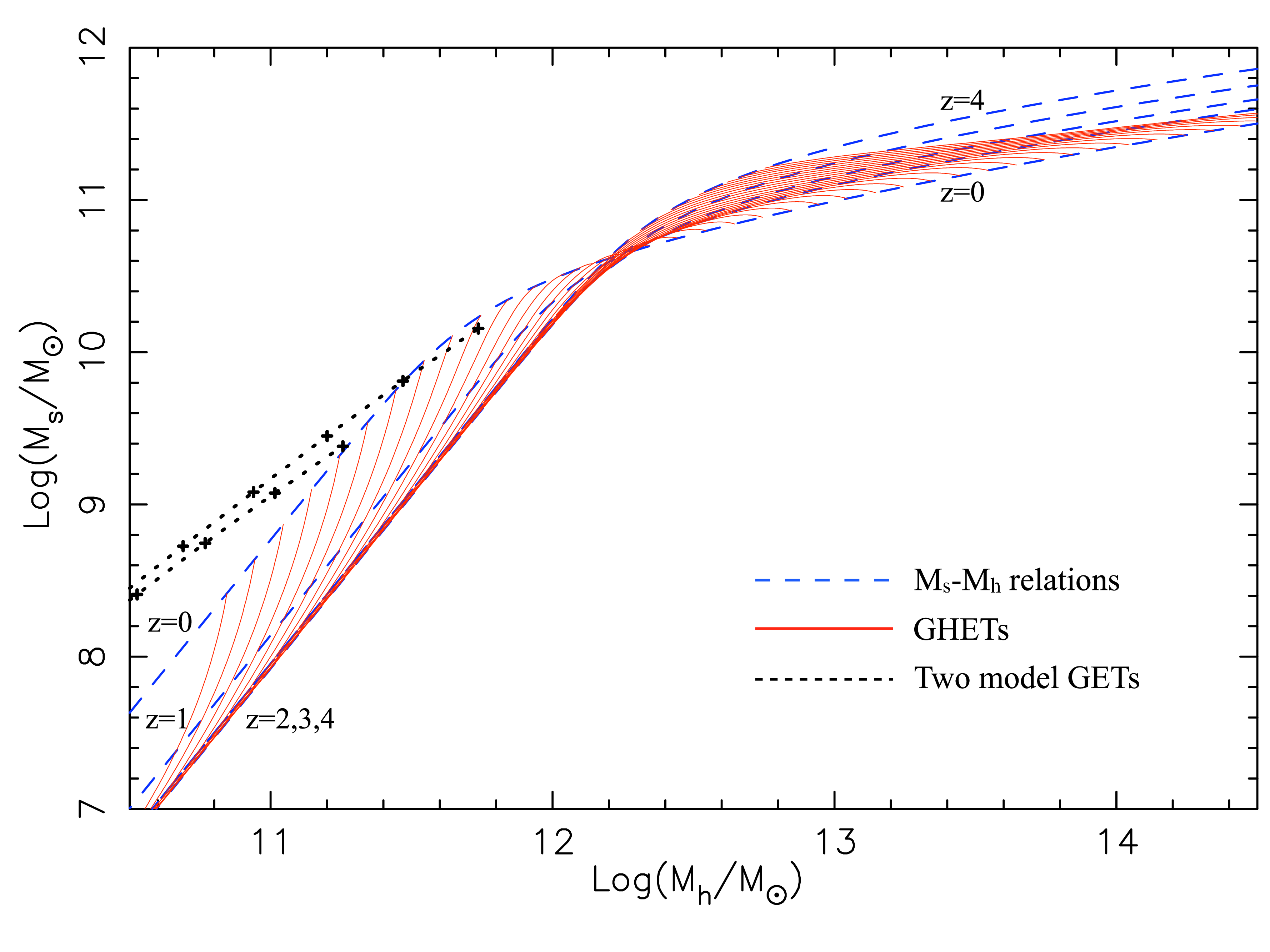} 
\caption{The \ms--\mh\ relations (blue dashed lines) at $z=0$, 1, 2, 3, and 4, from top to bottom 
(bottom to top) in the left (right) side, respectively, and the inferred average GHETs displayed
from $z=4$ to $z=0$ (red thin solid lines). 
The lower and upper dotted curves are model predictions for the evolution of two disk galaxies that 
end today with log(\ms/\msun) = 9.40 and 10.15, respectively (see \S 4 for details).
From right to left, the crosses correspond to $z= 0, 1, 2, 3,$ and 4.
}
\label{MsMh}
\end{figure*}

\subsection{The dark halo mass aggregation history (MAH)}

The \ms(\mh,$z$) relationship proposed in the previous section allows us to 
transform an average dark halo MAH, \mh($z$), into an average stellar mass growth
history, \ms(\mh($z$),$z$) (the GHET). Instead of using a parameterization 
for the average MAHs as was done in CW09, here we calculate them by using the special
Extended Press-Schechter formalism developed in Avila-Reese, Firmani \& Hern\'andez 
(1998, see also Firmani \& Avila-Reese 2000). Under the assumption that the primordial density field is 
Gaussian, we calculate the overall mass distribution of progenitor halos at a given 
epoch $z_{i+1}$ that will be contained in a halo of mass $M_i$ at a later epoch $z_i$. 
Thus, after defining the cosmology and the mass power spectrum of fluctuations (as in BCW10, 
parameters very close to WMAP5 are used here, Komatsu et al. 2009), we start from a given 
mass at a given time (e.g., $z=0$) and calculate its mass distribution of progenitors 
at a previous time step through Monte Carlo trials. The mass of the most massive halo 
(main progenitor) is used as the conditional for the next time step and so on. 
This way is obtained one realization of the MAH. For each \mh\ given at $z=0$, we calculate 
20000 random tracks (MAHs) and then we average among them to get the average MAH for 
this mass. In spite of the stochasticity, the MAHs present a systematical (hierarchical) 
trend with mass: the smaller is \mh($z=0$), the earlier is its mass assembly on average
(Avila-Reese et al. 1998; van den Bosch 2002; Wechsler et al. 2002). 
Our average MAHs are not easy to describe by a simple parametrization, but they are in 
good agreement with those measured in the outcome of cosmological N-body simulations 
(e.g., Fakhouri, Ma, \& Boylan-Kolchin 2010).

\begin{figure*} 
\vspace{11.5cm}
\includegraphics{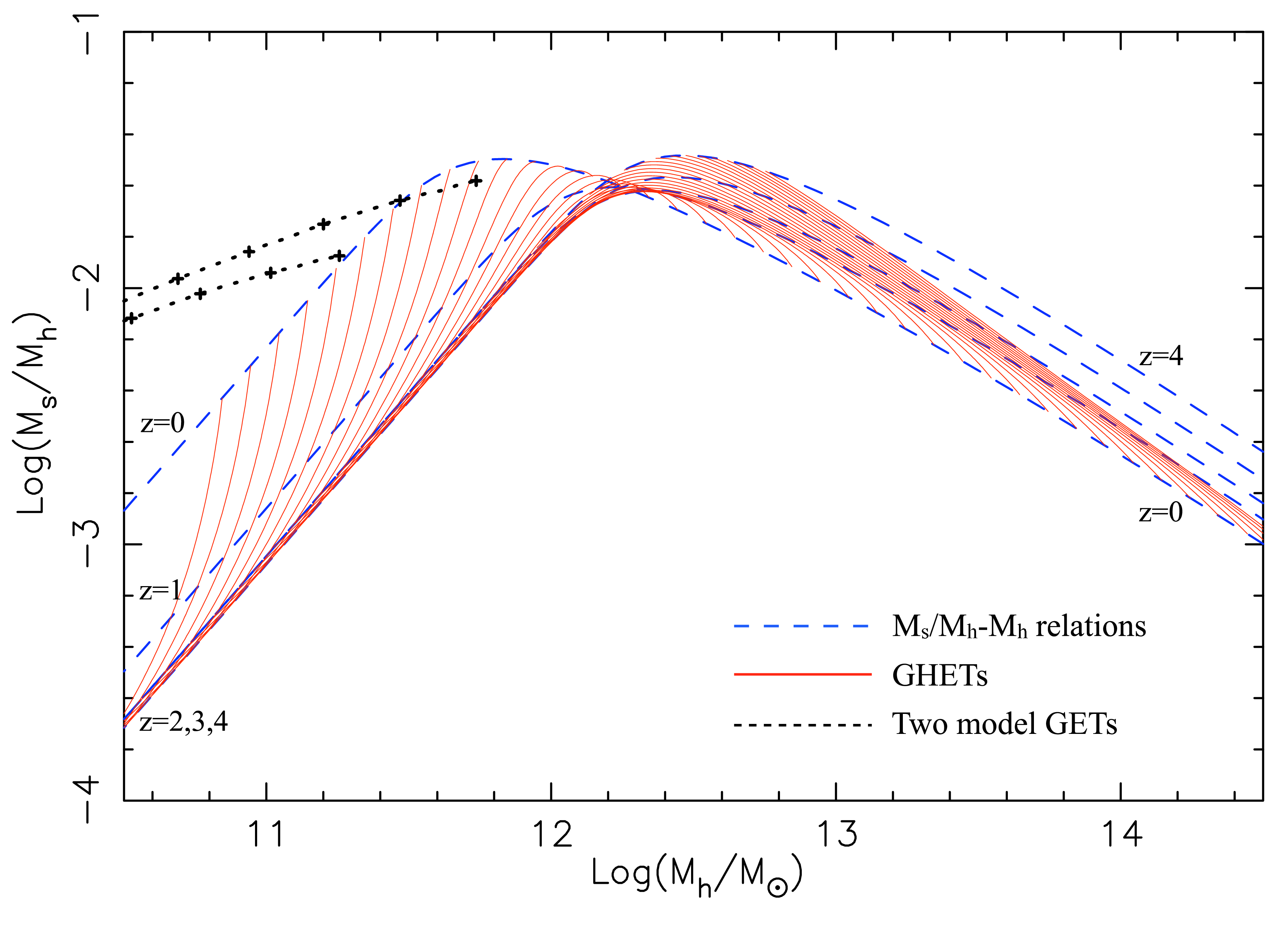} 
\caption{Same as in Fig. \ref{MsMh} but for the \ms-to-\mh\ ratio.
}
\label{galefficiency}
\end{figure*}

\begin{figure}
\vspace{7.6cm}
\includegraphics{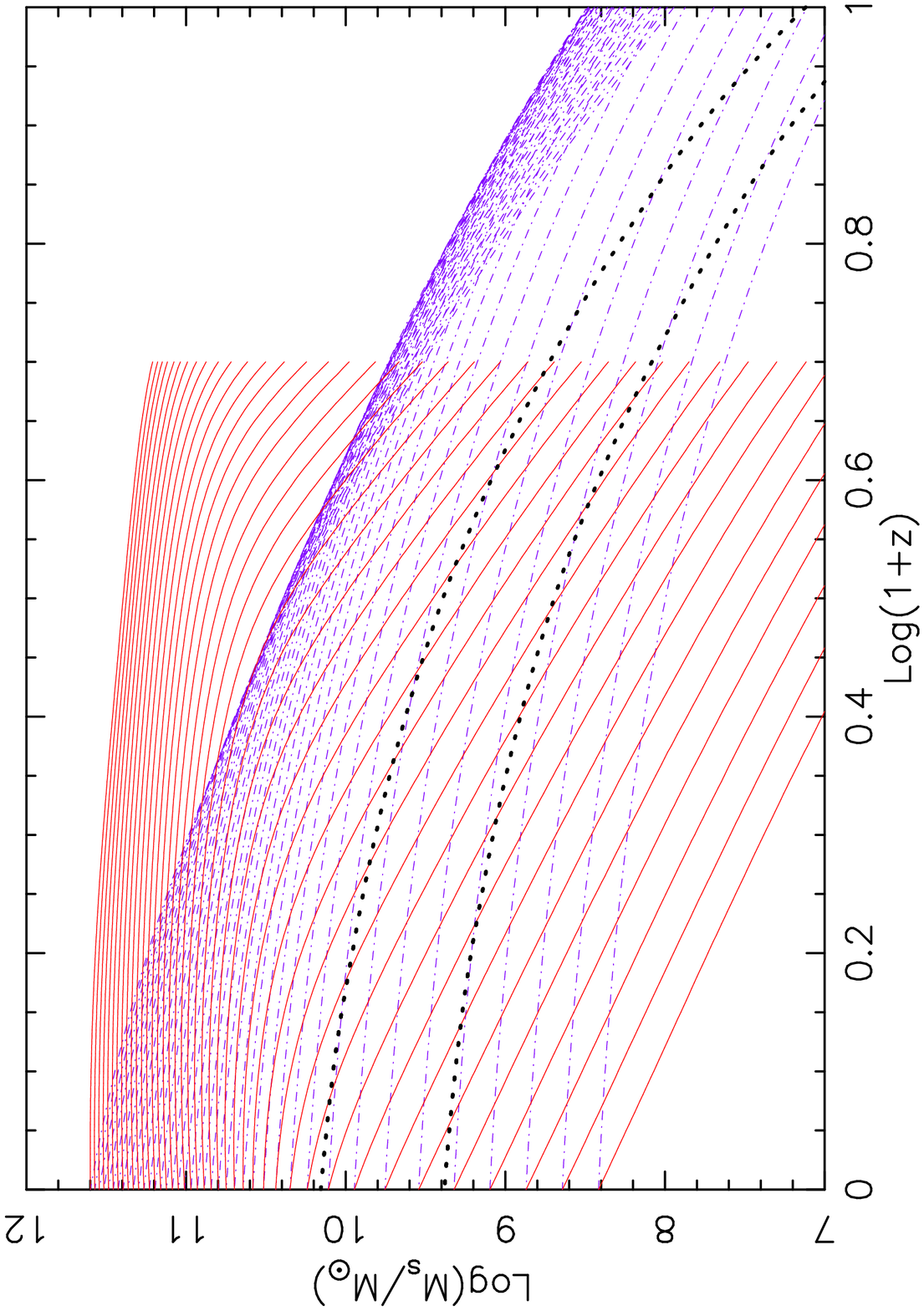}
\caption{Average GHETs (red solid lines) and the corresponding halo average MAHs (violet 
dot-dashed lines); for comparison purposes, each one of the latter ones was shifted vertically 
in such a way that \mh($z$=0) = \ms($z$=0). The lower and upper dotted curves are model predictions 
for the evolution of two disk galaxies that end today with log(\ms/\msun) = 9.40 and 10.15, 
respectively (see \S 4 for details).
}
\label{GHETs}
\end{figure}

\section{Galaxy stellar mass buildup: average evolutionary tracks}

The GHETs in the \ms\ vs. \mh\ diagram, calculated as explained in \S 2, 
are plotted with red thin solid lines in Figure \ref{MsMh}. 
The blue dashed curves show the \ms--\mh\ relations for $z=0, 1, 2, 3, 4$ 
from top to bottom (bottom to top) on the left (right) side. The GHETs in 
Figure \ref{MsMh} start from the $z=4$ relation and end at the $z=0$ one. 
Two features of the GHETs in such a diagram are remarkable. 
(1) Below log(\mh/\msun)$\sim 11.6$ the average slopes of the GHETs are 
$d$log\ms/$d$log\mh$\sim 4.5$, greater than the slopes $d$log\ms/$d$log\mh$\sim 2.3$ 
of the \ms--\mh\ relations in their low-mass side at any $z$.  
(2) The GHETs of massive halos (log(\mh/\msun)$\grtsim 12$ at $z=0$) stop to 
grow, happening this earlier, the more massive are the galaxies. 
Such a behavior is due to the folding seen in the \ms--\mh-$z$ surface that 
inverts the spatial distribution of the \ms(\mh,$z$) curves  going from the 
left to the right side of the diagram.  The same properties may be derived 
from Figure \ref{galefficiency}, where the GHETs are shown as tracks
that connect the \ms/\mh--\mh\ relations. Here the folding is seen in the blue 
dashed lines as a maximum that shifts to the higher--mass side as $z$ 
increases, making the lines cross each other. 
Both figures help to understand why the GHETs of massive halos attain a maximum. 

The GHETs ending at $z=0$ with masses $8.4 \le $log(\ms/\msun) $\le 11.6$ 
are plotted with red solid lines in Figure \ref{GHETs}. The corresponding halo 
average MAHs are also plotted but, for comparative reasons, each one has been 
shifted vertically in such a way that each MAH coincides with its related
GHET at $z=0$ (blue dashed lines). This figure shows in detail the difference 
between the average stellar and the halo evolutionary track shapes. 
For \ms($z$=0)$ < 10^{10.5}$ \msun, each GHET grows faster than the 
corresponding MAH since $z=4$. For \ms($z$=0)$ > 10^{10.5}$ \msun, as the system 
is more massive, the stellar assembly of the galaxy occurs earlier in time with 
respect to the corresponding halo. Our analysis shows conclusively that the 
stellar mass growth of galaxies deviates from the halo mass growth, 
specially for low- and high-mass systems.

Figures \ref{MsMh} and \ref{galefficiency} show that the more massive 
the galaxies, the earlier their GHETs start to slow down their \ms\ growth
and at each epoch there is a typical halo or galaxy mass
at which this growth stagnates; this typical mass decreases with time. 
Such a situation can be interpreted as a transition at each epoch from active 
to passive galaxy populations, a transition at early epochs happens only 
for the largest galaxies but at later epochs involves gradually 
smaller galaxies (see \S\S 3.2 for a more quantitative discussion). 

This transition from active to passive regimes is observed in the evolution of 
the blue and red components of the \gsmf. 
In fact, observations show directly that, despite the stellar mass growth, the blue (late-type) 
galaxy component of the \gsmf\ is not enriched with massive galaxies with time; on the contrary, 
within the uncertainty, it seems to become poorer in massive galaxies (e.g., Bundy et al. 2006; 
Bell et al. 2007; Pozzetti et al. 2010). Such a phenomenon is explained by a migration
of galaxies from the blue to the red component of the \gsmf; the red (early-type) galaxy 
component of the \gsmf\ is continuously fed with the migrated galaxies. Our result, in the 
light of such an interpretation, is remarkable because in our analysis the galaxy color 
bi-modality never has been taken into account. Studies where the \gsmfs\ were decomposed 
into blue and red populations support our result in an independent way 
(e.g., Bell et al. 2007; Drory et al. 2009; Pozzetti et al. 2010).

Note that the prediction of a population transition and the typical mass that 
separates the two populations at each $z$ depends on the position of the folding 
in the \ms--\mh--$z$ surface. This feature reveals a fundamental aspect of the \gsmf\ 
evolution. Therefore, the accurate study of this feature deserves maximum efforts 
from the observational point of view. 

The fact that the more massive are the systems, the earlier they finish their 
\ms\ assembling, transiting from the active to the passive population, will be called here 
{\it population downsizing} (in the literature the term 'archaeological downsizing' 
has been used to describe a related behavior of galaxies; see e.g., Fontanot et al. 
2009, and the references therein).

For systems less massive than $\ms\approx 10^{10.5}$ \msun\ at $z=0$ 
($\mh\approx 10^{12}$ \msun), the active (growing) phase continues still at $z=0$ 
on average (the population downsizing has not happened), and the smaller is the system, 
the \ms\ growth is more delayed and concentrated towards the present.
The fact that the less massive is the system, the later happens its active phase 
of \ms\ assembling, is called {\it downsizing in SSFR} (see e.g., Fontanot et al. 
2009, Firmani et al. 2010 and the references therein).

In the different diagrams plotted in Figures \ref{MsMh}, \ref{galefficiency}, and \ref{GHETs}, we have 
introduced the corresponding evolutionary tracks (crosses at $z=0, 1, 2, 3,$ and 4 connected with 
dotted lines) for two (low--mass) galaxies calculated by means of a \lcdm--based model of galaxy evolution 
which will be explained in detail in \S 4. Their final stellar masses are 
log(\ms/\msun) = 9.4 and 10.1, respectively. We anticipate the rather different behavior 
of the model 'galaxian evolutionary tracks' (hereafter GETs) with respect to the GHETs. 
For example, the average slope of the former in the \ms--\mh\ diagram is $d$log\ms/$d$log\mh=1.4, 
much less than the slopes of any GHET, and also less than the lower limit estimated above 
for the low--mass side of the \ms--\mh\ relations at each $z$. This means that the model isochrones 
(defined by connecting crosses of a given $z$) have a distribution and evolution opposite to 
that of the empirical \ms--\mh\ relations. 

\begin{figure}
\vspace{7.6cm}
\includegraphics{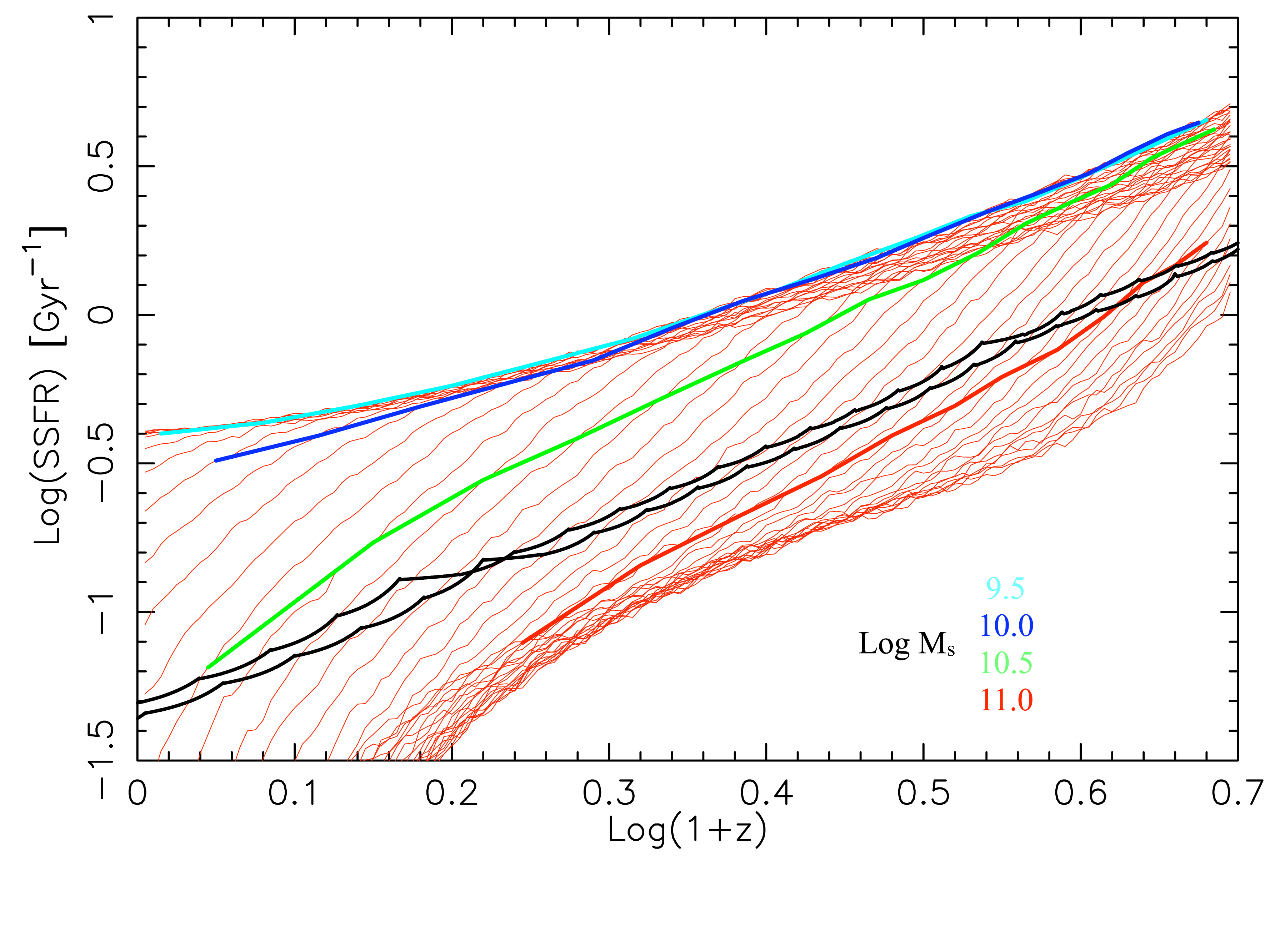} 
\caption{Evolution of the SSFR (see text for its definition) corresponding to the same GHETs shown 
in Figure \ref{GHETs} (solid red lines).  
From top to bottom, the curves with thick solid lines connect the tracks at the moment log(\ms/\msun) 
equal to 9.5 (blue), 10 (cyan), 10.5 (green), and 11.0 (red), respectively.
The upper and lower black curves are model predictions for the evolution of two disk galaxies that 
end today with log(\ms/\msun) = 9.4 and 10.1, respectively (see \S 4 for details).}
\label{dotM}
\end{figure}

\subsection{Stellar mass growth rates as a function of mass}

The stellar mass buildup of galaxies may happen due to in situ SF or accretion of stars,
mainly in dry merger events. Several theoretical (e.g., Maller et al. 2006; Guo \& White 2008), 
semi-empirical (Zheng, Coil \ Zehavi 2007; CW09; Yang, Mo \& van den Bosch 2009; 
Wang \& Jing 2010), 
and observational 
(e.g., Bundy et al. 2006,2009; Bell et al. 2007; Drory \& Alvarez 2008) pieces 
of evidence show that the former channel completely dominates in low- and intermediate-mass galaxies
at all epochs, while the latter may play a moderate role for massive 
($\ms\grtsim 10^{11}\msun$, mainly red) galaxies at later epochs ($z\lesssim 1$). 

As a working hypothesis, we will assume that the stellar mass buildup implied by our derived 
average GHETs is only due to in situ SF.
Therefore, we calculate the SSFR as the time derivative of \ms($z$) divided by the current mass, 
\dotMz, and divided by ($1-R$), where $R=0.4$ is the gas recycling factor due to stellar mass 
loss. The obtained SSFRs are then compared with directly observed SSFRs as a function of 
\ms\ and $z$.  

In Figure \ref{dotM}, the evolution of the GHET-based SSFR is plotted (red solid line) for 
the same cases shown in Figure \ref{GHETs}. From top to bottom, the mass of the GHETs 
increases, except for the minor inversion (track crossing) present in the top right side 
of the diagram. For all the masses, SSFR decreases as $z$ decreases.
The overall behavior of decreasing SSFR with time agrees, by construction, with the known direct 
observational inferences of the cosmic stellar mass density history and its time derivative 
(e.g., Drory et al. 2005; P\'erez-Gonz\'alez et al. 2008), providing a comparison term to the observed 
cosmic SFR density history (Madau et al. 1996; see Hopkins \& Beacom 2006 for an extensive compilation
of observational inferences). 

\begin{figure}
\vspace{8.8cm}
\includegraphics{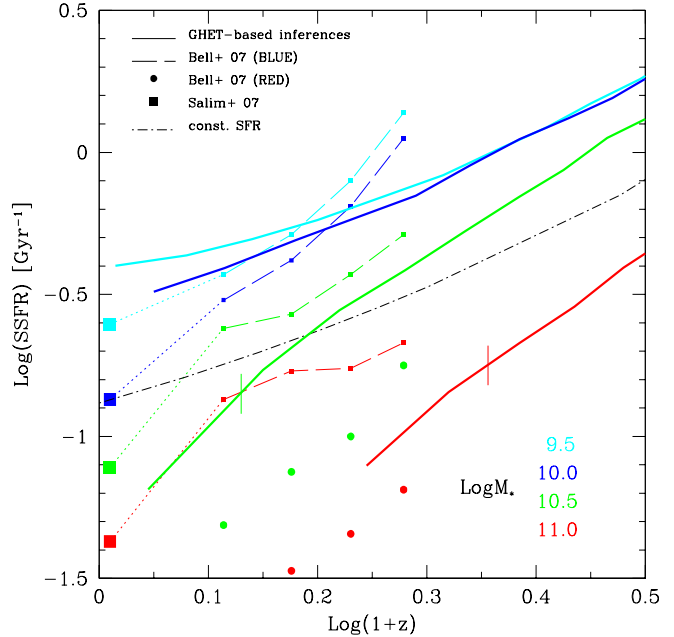}
\caption{Same curves connecting the GHETs at four values of  \ms=constant (indicated inside the panel) 
as shown in Figure \ref{dotM} (solid thick lines). 
The filled squares at $z\sim 0$ are estimates from SDSS by Salim et al. (2007), and the small squares 
connected by dashed lines were inferred from Bell et al. (2007) for their sample of blue galaxies; 
in both cases the (average) masses are the same as the GHET-based masses. 
The filled circles are also inferences from Bell et al. but for their sample of red galaxies and only for 
log(\ms/\msun)= 11.0 (red filled circles) and 10.5 (green filled circles). The vertical marks over
the log(\ms/\msun)= 11.0 and 10.5 curves indicate the typical redshift at which these masses transit from
active to passive phases.
Note that the end of the $10^{11}$ \msun\ GHET--based curve (when these masses are already in the 
passive phase) agrees well with the SSFR of red rather 
than blue galaxies; the end of the $10^{10.5}$ \msun\ GHET--based curve is in between the red and blue 
galaxy samples. 
The dot--dashed line shows the curve $1/[t_H$($z$) -- 1 Gyr]($1-R$) corresponding to a constant SFR. 
}
\label{dotMobs}
\end{figure}

We test now if the GHET-based SSFRs histories as a function of mass are in agreement with 
the direct observational measures. 
The observations of course do not refer to individual evolutionary tracks.
What observers do is to determine the SSFRs of galaxy samples in different redshift bins. 
A commonly reported result is the average of the SSFR in different $z$ bins corresponding to a given 
range of \ms\ (the same at all $z$'s). 
In Figures \ref{dotM} and \ref{dotMobs}, from top to down, the thick (cyan, blue, green, and red) 
lines correspond to the SSFR inferred from the GHETs that in each $z$ have log(\ms/\msun)=9.5, 
10.0, 10.5, and 11.0, respectively. 

In Figure \ref{dotMobs}, we reproduce averages of SSFR observational estimates at different $z$'s for 
galaxies in four \ms\ bins (centered in the same masses plotted for the GHETs). 
These estimates were taken for $z\sim 0$ from Salim et al. (2007) excluding galaxies with AGN, and 
for higher $z$'s (until $z\sim 1$), from Bell et al. (2007; only their Chandra Deep Field South 
estimates are plotted here). 
The data used to get these estimates refer to the population of blue (star-forming) galaxies 
(see Firmani et al. 2010 for more details on the choice of the observational data). 
It should be stressed that the intrinsic spread around the plotted curves is rather large.

For the largest mass bin (\ms\ about $10^{11}$ \msun, lower red curves), the GHETs imply clearly 
lower SSFRs than the directly measured SSFRs of star--forming (blue) galaxies at $z\lesssim 1$. 
This is expected since, as discussed above, the most massive galaxies show evidence to have early 
transited to the passive population. For the galaxy mass $\ms\approx 10^{11}\msun$, the transit to 
the passive, red sequence happens at $z\approx 1.3$ (see Fig. \ref{Mtran} below) as indicated 
by a vertical mark over the curve. Bell et al. (2007) reported the estimated SSFRs also for 
the red sequence galaxies. Their estimates for $\ms\sim 10^{11}$ \msun\ (red circles) and 
$10^{10.5}$ \msun\ (green circles) are plotted in Figure \ref{dotMobs}. The directly measured 
SSFRs of massive (red) galaxies are very low, decreasing their values as $z$ decreases in the 
same manner as our predictions show. While 
speculative by the moment due to the large uncertainties, the fact that the values of our inferred 
SSFRs are systematically higher than the direct measures of red massive galaxies could be 
interpreted as an evidence of contribution of stellar accretion (dry mergers) to the stellar mass 
growth of these  galaxies at late epochs (see Section 2.1 for a discussion on this possibility).

Recently, Damen et al. (2009a) presented estimates of the SSFR up to $z\approx 2.8$ 
for massive galaxies ($\ms> 10^{10.5}$ \msun) in the same Chandra Deep Field South
used in Bell et al. (2007). As mentioned above, for redshifts $z\grtsim 1$ galaxies of
$\ms\approx 10^{11}$ \msun\ or smaller are on average in their active (SF-driven) growth 
regime. The results by Damen et al. are in rough agreement at these redshifts with our 
predictions. If any, the GHET-based SSFRs are slightly lower than the averages
of the directly measured SSFRs.

For the intermediate mass bin (\ms$\approx 10^{10.5}$ \msun, green curves), the agreement between 
the average SSFRs inferred from the GHETs and those measured directly for the star-forming (blue) 
galaxies (Bell et al. 2007) is satisfactory, given the spread of data in SSFR. 
If any, the SSFR from the GHETs is slightly lower, in particular at low $z'$s. 
Again, due to the population downsizing, active (blue) galaxies of masses \ms$\approx 10^{10.5}$ \msun\ 
transit to the passive (red) population at $z\approx 0.35$ (see Fig. \ref{Mtran}) as indicated
by the mark over the green curve, so that their SSFRs at lower $z$'s should be in between
those measured for the blue and red sequences. The agreement in the mass range around  
$10^{10.5}$ \msun\ with the Damen et al. (2009a) observational inferences up to $z\approx 2.8$ 
is also good, though the samples of these authors do not separate galaxies in early- and late-type 
ones. For $z> 0.5$ the slope of the SSFR(\ms$\approx 10^{10.5}$\msun)--($1+z$) relation 
for our prediction and for direct observations is about 3.3. 

For the lowest mass bins ($\ms\sim 10^{10.0}$ and $10^{9.5}$ \msun, blue and cyan curves), 
the agreement between GHET-predicted and directly-measured SSFRs is reasonable if takes 
into account (i) the large intrinsic spread of the observational inferences 
(1$\sigma$ width of $\sim 0.5$ dex around the mean in the relation log(SSFR)--log\ms\ 
at $z\sim 0$, Salim et al. 2007; for higher redshifts, the typical reported intrinsic 
1$\sigma$ widths are of $\sim 0.2-0.3$ dex, Noeske et al.2007; Zheng et al. 2007; 
Damen et al. 2009a,b); and (ii) the sample incompleteness, which increases
as $z$ is higher (for such small masses, this may bias the observational average 
estimates toward larger values for $z>0.3-0.5$). If any, the predicted 
SSFR(\ms=const)--($1+z$) curves for these low masses decrease slower (are shallower) 
than the curves inferred from direct observations for $z<1$ (see also Damen et al. 2009a). 
Better observational estimates for both the low-mass \gsmf\ and SSFRs at high 
redshifts are needed to attain more conclusive results. 

The dot-dashed line in figure \ref{dotMobs} shows the curve 
$1/[t_H$($z$) -- 1 Gyr]($1-R$) corresponding to a constant SFR. 
Galaxies above (below) this curve are currently forming stars at a 
rate higher (lower) than the past average.

\begin{figure}
\vspace{7.6cm}
\includegraphics{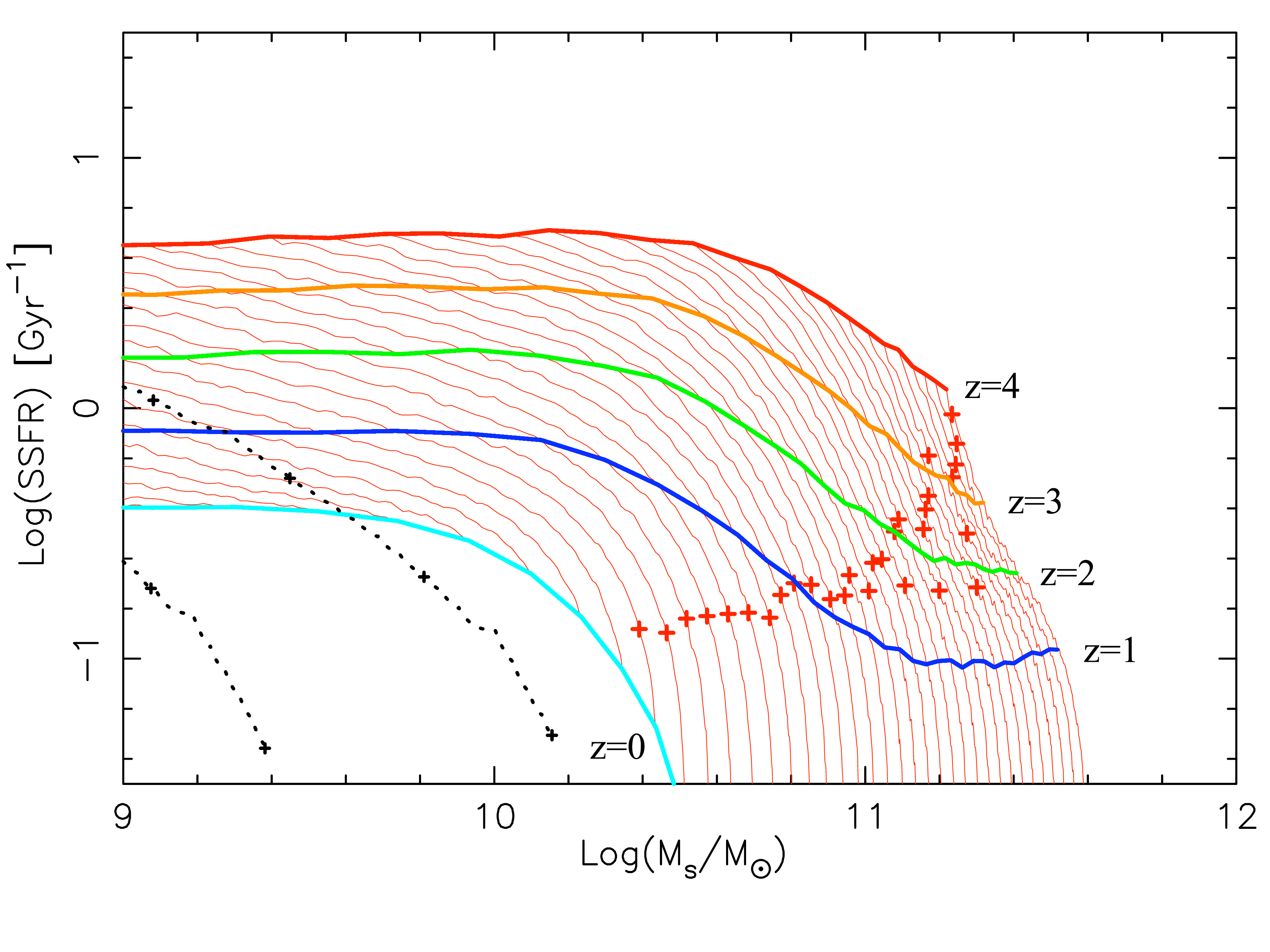} 
\caption{Same GHETs as shown in Figure \ref{GHETs} but in the SSFR vs \ms($z$) diagram (thin solid red lines). 
From bottom to top, the curves with thick solid lines connect the tracks at a given epoch (isochrones): $z=$0  
(cyan), 1 (blue), 2 (green), 3 (orange), and 4 (red), respectively. 
The red crosses indicate the points where the GHET slope declines below -5.
The lower and upper dotted curves are model predictions for the evolution of two disk galaxies that end today 
with log(\ms/\msun) = 9.40 and 10.15, respectively (see \S 4 for details).}
\label{SSFR-Ms}
\end{figure}

In Figure \ref{SSFR-Ms}, the evolution of SSFR vs. \ms\ for the same GHETs of previous figures 
is plotted (red solid lines). This plot remarks the fact that small galaxies keep high 
SSFRs and grow fast at late epochs, while very large galaxies had high SSFR and rapid 
growth in the remote past, being passive at later epochs.
In Figure \ref{SSFR-Ms}, from bottom to top the thick (cyan, blue, green, orange, red) 
lines connect the individual tracks at the redshifts $z= 0, 1, 2, 3,$ and 4 (isochrones), respectively. 
Each one of these isochrones corresponds to the average SSFR--\ms\ relations at a given $z$ 
and can be compared with direct observational inferences of SSFR vs \ms\ at different $z$ bins. 
These inferences for samples at different epochs (from $z\sim 0$ up to $z\sim 2$)
show that galaxies (mainly star-forming ones) are ordered in the logSSFR--log\ms\ 
diagram along sequences of intrinsic widths 1$\sigma\lesssim 0.3-0.5$ dex (e.g., Bauer et al. 2005;
Salim et al. 2007; Schiminovich et al. 2007; Noeske et al. 2007; Bell et al. 2007; 
Drory \& Alvarez 2008; Damen et al. 2009a,b; Santini et al. 2009; Oliver et al. 2010; 
Rodighiero et al. 2010). The averages of these sequences are such that the smaller is 
\ms, the higher the SSFR. However, due to the sample incompleteness limit, which increases 
from a few $10^9$ \msun\ to several $10^{10}$ \msun\ from $z\sim 0$ to 1, respectively, these
averages at low masses might be overestimated. The isochrones showed in Figure 
\ref{SSFR-Ms} are in general within the 1 $\sigma$ width of intrinsic spread of the observational
SSFR-\ms\ relations. 
At low masses (where observational samples with measured SFRs are below the completeness limit), the 
isochrones tend to be below the direct observational estimates of SSFR at the given epoch but at 
intermediate masses the agreement is rather good, i.e., at low masses, the GHET-based SSFR-\ms\ 
relations tend to be shallower than those constructed from direct observations.

At the high mass end, our isochrones fall faster than the averages of the observed SSFR--\ms\ sequences,
which typically are reported for star-forming (blue) galaxies. Have in mind that our results refer to {\it average} 
trends and in this sense the average high mass galaxies in our case already transited to the passive 
(red) population. In fact, the SSFR--\ms\ relation at a given $z$ should be constructed separately
for active (blue) and passive (red) galaxies. Therefore, in the SSFR--\ms\ diagram two sequences would appear: 
at low masses a sequence dominated by blue galaxies and at high masses the sequence dominated
by red galaxies. From the point of view of the GHET analysis, the characteristic mass that divides 
both sequences is related to the transition mass inferred from the GHETs at each $z$ (see \S\S 3.2 below). 
As seen in Fig. 7, the average slope of the passive sequence (high-mass side) at any $z$ is steeper 
than that one of the active sequence (low-mass end). Interestingly enough, this is the trend actually 
revealed in those observational works where the SSFR--\ms\ relations at different $z'$s (out to 
$z\sim 1.0-1.5$) are plotted separately for red and blue galaxies (Bell et al. 2007; Oliver et al. 2010).

In general, we conclude that the SSFRs at different $z$'s and as a function of \ms\ predicted on the
basis of our GHETs are reasonably consistent with recent observational studies based on direct measures 
of the SSFR of galaxies at different $z$'s (especially those that divide galaxies into blue and red ones). 
This consistency is particularly important {\it because the observational information concerning SFR never
enters in our scheme}. Therefore, on the one hand, this result is an independent test for the whole 
approach presented here (it predicts the mean SSFRs of galaxies at different epochs consistent 
with direct observational estimates). On the other hand, it supports our hypothesis that the 
stellar mass buildup given by the average GHETs 
is driven by in situ SF. Neither does this mean that the growth of \ms\ by accretion of stellar systems 
(dry mergers) may not happen, but it suggests that it is less dominant than the SF channel (our approach 
and the accuracy of current estimates of SSFR at different redshifts do not allow for quantitative 
inferences of the contributions of one or another channel of galaxy mass growth). 

A natural question arises about the information the GHETs provide on the stellar initial mass function
(IMF). We have outlined the agreement between our SSFR derived from the stellar mass growth and 
the directly inferred SSFR. The latter inference is highly sensitive to the assumption of a
universal IMF. In this sense, we can say that such an agreement is congruent with a universal IMF.
Nevertheless, the uncertainties are yet large, making it premature to draw such a conclusion as 
sufficiently robust.

Finally, the results mentioned above confirm and integrate into a unified picture two key facts of 
galaxy stellar mass assembly: (1) the downsizing in SSFR of low--mass galaxies and 
(2) the population downsizing of massive galaxies. The former will be discussed on the light 
of \lcdm-based models in \S 4. The latter is quantified in the next section.

\subsection{Evolution of the characteristic mass that separates active from passive population}

\begin{figure}
\vspace{6.5 cm}
\includegraphics{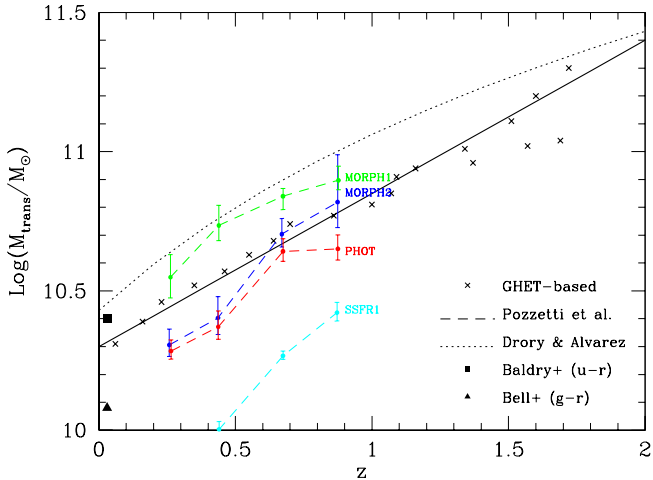}
\caption{Transition mass as a function of $z$ (diagonal crosses), defined as the mass corresponding 
to the GHET that at a given $z$ started to brake its \ms\ growth (the slope of the SSFR--\ms\ relation 
becomes steeper than $-5$). 
Masses above \mtran($z$) are already in the passive phase (red quiescent population).
The solid line is the dependence given in eq. (\ref{Mtranfit}).
The filled circles with error bars connected by lines are the crossing masses, M$_{\rm cross}$, 
at different $z$'s determined in Pozzetti et al. (2010) as the cross of the \gsmfs\ of 
the early-- and late--type populations. The labels at the end of each curve indicate 
the different schemes adopted by the authors for the classification: MORPH1 and MORPH2 
are morphological classifications using their {\it ZH} and {\it MRS} classification schemes, 
respectively; PHOT refers to a photometric classification scheme; SSFR1 
corresponds to a classification into active and quiescent galaxies according to their SSFR.
The dotted line is the law given in Drory \& Alvarez (2008) for the mass above which the SFR as 
a function of mass begins to drop exponentially. 
The filled square and triangle are local determinations of the crossing mass of local blue and red 
\gsmfs\ by Baldry et al. (2004) and Bell et al. (2003), respectively.
}
\label{Mtran}
\end{figure}

Our method reveals the existence of a characteristic stellar mass at each 
$z$, \mtran($z$), at which the average GHET sharply declines its growth rate.
We interpret this as a transition from the active to the quiescent regime of SFR, 
the onset of the migration from the star-forming blue to the passive red population 
(population downsizing). We may estimate quantitatively \mtran($z$) by identifying the mass 
at a given $z$ corresponding to the GHET that started to strongly decrease its SSFR, for 
example, when the slope of the track in SSFR vs. \ms\ (see crosses in Figure \ref{SSFR-Ms}) 
becomes steeper than $-5$ (the measured SSFR at this point increases from 
log(SSFR/Gyr$^{-1}$)$ = -0.8$ at $z\sim 0$ to $\sim -0.5$ at $z\sim 3$).
The stellar \mtran\ calculated this way is plotted vs $z$ in Figure \ref{Mtran} (diagonal crosses). 
The function:
\begin{equation}
{\rm log}(\mtran/\msun) = 10.30 + 0.55 z
\label{Mtranfit}
\end{equation}
offers an approximate description of the results up to $z\sim 2$ (solid line).

The focus now is to compare our result with independent observational pieces of evidence for a 
transition mass from active to quiescent/passive population as a function of $z$. 
Nowadays, as mentioned in the Introduction, it is possible to follow the evolution of 
the early-- (red) and late--type (blue) \gsmfs\ separately.
In one of the most recent and complete works, based on the $z$COSMOS survey, Pozzetti et al. 
(2010) determined the mass where the early-- and late--type \gsmfs\ cross (different estimators 
for these two populations are used, see the figure caption) from $ z\approx 1$ to $z\approx 0.2$. 
Such a crossing mass, $M_{\rm cross}$, is interpreted namely as the typical mass of late-type 
galaxies migrating to early-type ones (see also Bell et al. 2007).

The agreement between the Pozzetti et al. (2010) results for $M_{\rm cross}$ (circles with error 
bars connected by different dashed lines) with our \mtran($z$) is satisfactory as can be seen 
in Figure \ref{Mtran}. Local estimates of $M_{\rm cross}$ by Bell et al. (2003; filled triangle) 
and Baldry et al. (2004; filled square) are also plotted, as well as the law inferred by 
Drory \& Alvarez (2008) for the mass above which the SFR as a function of \ms\ begins 
to drop exponentially, \mtran/\msun = $10^{10.43}$($1+z$)$^{2.1}$.
Our result also agrees qualitatively with other observational studies, with different 
definitions of the characteristic mass above which the passive (red) population of 
galaxies dominates in number density (e.g., Bundy et al. 2006; Hopkins et al. 2007; 
Vergani et al. 2008).

We may predict also the flow of galaxies transiting from active to passive populations 
at each $z$ per unit of co-moving volume. At each redshift interval $z, z+dz$ 
(time interval $t, t+dt$) there are GHETs transiting from the
active to the passive regime, whose associated halo masses are in the
interval $\mh, \mh+d$\mh. The transition rate in number density is then given
by the abundance of halos in such a mass interval divided by $dt$,
$\Phi(\mh,z)\times d$log\mh/$dt$. For the abundance of halos at each $z$ we use 
the $HMF$s given in Tinker et al. (2008) adapted to our cosmology and corrected 
to include sub-halos according to the functions given in BCW10. 

\begin{figure}
\vspace{7. cm}
\includegraphics{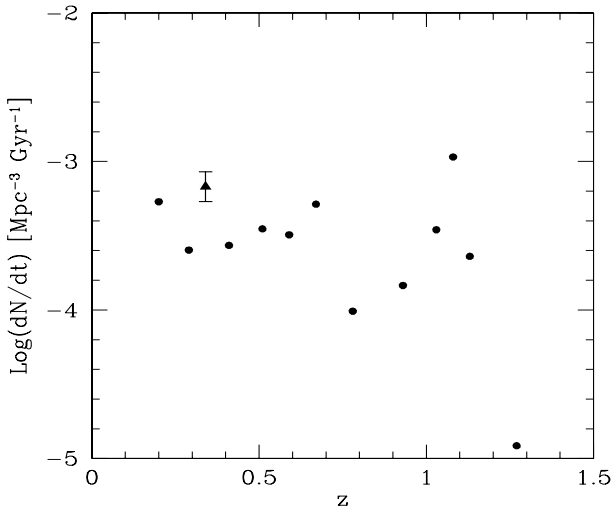}
\caption{Transition rate in number density of active to passive galaxies at each $z$.
The solid triangle with error bar is the estimate of growth rate in number density
of red galaxies reported in Pozzetti et al. (2010) integrated above log(\ms/\msun) = 9.8. 
}
\label{trans-rate}
\end{figure}

The result is shown in Fig. \ref{trans-rate}. The transition rate in comoving number 
density of active to passive galaxies up to $z=1$ scatters around $(1.0 - 5.5)\times 
10^{-4}$ gal Gyr$^{-1}$ Mpc$^{-3}$ without any clear trend with $z$ (the scatter is mainly 
due to the discretization of the MAHs in mass and $z$). At $z\approx 0.3$, the rate 
is $(3.7\pm 1.2)\times 10^{-4}$ gal Gyr$^{-1}$ Mpc$^{-3}$. The obtained passive population growth 
rates can be compared with estimates reported in the same work by Pozzetti et al. (2010). 
They found an average growth rate in number density of the red population integrated 
above log(\ms/\msun) = 9.8 of $6.8(\pm 1.2)\times 10^{-4}$ gal Gyr$^{-1}$ Mpc$^{-3}$ for a redshift
interval centered in $z=0.34$ (solid triangle with error bar in Fig. \ref{trans-rate});
this value is in reasonable agreement with our results. Note that the flow of galaxies 
in our case is related to a small mass range around the transition mass \mtran\ 
given by the {\it average} GHETs, while the direct observational determination takes 
into account a large range of masses. The fact that the latter
is only slightly larger than the former implies that indeed most of the galaxies
becoming red are those of masses close to the average transition mass.

\subsection{Reliability of the results}

The results presented above are based mainly on the evolution of the \ms--\mh\ relation.
An important question is then how general and reliable is the \ms(\mh,$z$) relationship
used here. A comprehensive analysis of the statistical and systematical uncertainties in 
\ms(\mh,$z$) and how they affect the shape of the \ms-\mh\ relations at different $z'$s 
has been carried out by BCW10. 
These authors separate the uncertainty sources into three classes: uncertainties 
(i) in the observational inference of \gsmf,  
(ii) in the dark matter $HMF$, and 
(iii) in the matching process arising primarily from the intrinsic scatter between \ms\ and \mh. 
They found that by far the largest contributor to the error budget of the local \ms--\mh\ relation 
comes from assumptions in converting galaxy luminosity into \ms, which amounts to uncertainties
of $\sim 0.25$ dex in the normalization of the relation (see the error bars in Fig.\ref{galefficiency-b}). 
The contribution from all other sources of error, including uncertainties in the cosmological model, 
is much smaller, ranging from 0.02 to 0.12 dex at $z=0$ and from 0.07 to 0.16 dex at $z=1$. 
At high redshifts ($z> 1$), statistical uncertainties grow, becoming as significant as the systematic ones.
On the other hand, the intrinsic scatter in \ms\ given \mh\ and the random statistical error in \ms\ 
inferences have a significant effect on the shape of the \ms--\mh\ relation at the massive end (taken 
into account in BCW10).

At $z\sim 0$, the results from most direct and indirect studies to infer \ms(\mh, $z\sim 0$), in spite 
of the different methods and different local \gsmfs\ used, agree well among them, especially if the 
uncertainties are taken into account (see BCW10 for a comparison).
The systematic shift of the (\ms/\mh)--\mh\ curves to higher \mh\ as $z$  increases 
(see Figs. \ref{galefficiency} and \ref{galefficiency-b}) is also a generic result obtained 
by all the authors (CW09; Wang \& Jing 2010; Moster et al. 2010; BCW10).
This implies that the trend reported here of downsizing in SSFR for galaxies less massive than 
$\ms\approx 3\ 10^{10} \msun$ is robust; the downsizing can be respectively stronger or weaker
depending on whether the low-mass end of the curves shifts more or less with $z$.
As the completeness of the samples at low masses improves, the estimate of the \gsmfs\ at their 
low-mass end will be better, and hence more accurately the evolution
of the \ms--\mh\ relation at lower masses will be constrained.

The results of different authors may differ regarding the exact location of the \ms--\mh\
(or \ms/\mh--\mh) curves at different $z$'s, especially for $\ms\grtsim 10^{10.5} \msun$.
For example, in Moster et al. (2010), the peaks in (\ms/\mh)--\mh\ significantly 
decrease with $z$ in such a way that the curves intersect only at very high masses.
This implies that the GHETs keep growing until $z=0$ even for large galaxies, the transition 
to passive population happening only for the most massive ones.
In a similar way, CW09 concluded from their analysis that out to $z\sim 1$ and for 
$\ms < 10^{11}$ \msun\ there is not a characteristic mass at which the growth rate 
of galaxies is truncated. However, this is in conflict with independent observations 
that show instead strong population downsizing (see \S\S 3.2 and Fig. \ref{Mtran}).

In the case of BCW10, the peaks in (\ms/\mh)--\mh\ slightly decrease out to $z\sim 1$ 
and for higher $z'$s, increase attaining at $z\approx 4$ the same level as at $z\approx 0$
(Figs. \ref{galefficiency-b} and \ref{galefficiency}). 
However, the uncertainties in the determination of the (\ms/\mh)--\mh\ curves are large and other 
behaviors with $z$ are allowed. The relationship used here closely resembles the
BCW10 results as discussed in \S 2. 

In order to probe the robustness of the conclusions presented here, we have performed our analysis 
by using \ms(\mh,$z$) relationships different to the one assumed in \S 2 but yet within the 
uncertainties of the BCW10 inferences. For example, we explored an \ms(\mh,$z$) relationship
such that the peaks of the (\ms/\mh)--\mh\ curves remain at the same level for all redshifts (i.e., 
$\chi$($z$)$ = 0$ in Eq. 2). The obtained GHETs are qualitatively similar to the ones presented here
and Eq.(\ref{Mtranfit}) slightly changes to ${\rm log}\left(\mtran/\msun\right) = 10.2 + 0.6 z$.
It is noteworthy that the high mass GHETs in Fig. \ref{GHETs}, presenting a 
stagnation in their \ms\ growth, show a shape a little sensitive to the parameter $\delta_a$. 
By assuming $\delta_a = 0$ in Eq. (2), the most massive GHETs, after reaching a maximum stellar 
mass at approximately the same $z$ as presented in Fig. \ref{GHETs}, decline in mass but not by 
more than 0.1 dex. 

We conclude that the main results reported here are robust at least for the 
range of variations in the \ms(\mh,$z$) relationship that remains within the
current uncertainties in the inference of this relationship.

\section{Model predictions versus hybrid evolutionary tracks}

Any model of galaxy formation and evolution within the hierarchical \lcdm\ scenario should predict 
how does the \ms-to-\mh\ ratio (the ``galaxy formation efficiency'') of individual galaxies evolve. 
Several complex astrophysical processes intervene in shaping the evolution of this ratio.
Most of the semi-analytical and semi-numerical models --as well as the numerical simulations in that 
regards subgrid physics-- introduce physical schemes for modeling these processes and for reproducing 
the {\it present--day} \ms--\mh\ relation (or the luminosity function) and other properties/correlations 
of galaxies.
The question now is whether such models are able to reproduce the whole evolution outlined here.
Different approaches to explain the population downsizing have been recently proposed,
while the SSFR downsizing is just starting to be recognized as a problem
(e.g., Fontanot et al. 2009; Firmani et al. 2010; Col\'{\i}n et al. 2010).   

In each one of the figures where the GHETs were presented in this paper, we have plotted the
GETs corresponding to two low--mass disk galaxy models that end today with log(\ms/\msun) = 9.40 
and 10.15  (log(\mh/\msun) = 11.25 and 11.73; dotted lines).
The models were calculated with a semi-numerical method that self-consistently solves dynamical 
and hydrodynamical equations of: halo mass virialization given the MAH, disk formation in centrifugal 
equilibrium inside the growing halo, gravitational drag of the disk over the halo inner regions, 
SF triggered by gravitational instability and self-regulated by a vertical energy balance between 
energy input due to SNe and turbulent energy dissipation, and SN--driven mass outflows 
(Firmani \& Avila-Reese 2000; Firmani et al. 2010).
The main assumptions implicit in the models are those of: spherical and cylindrical symmetries for the 
halo and disk, respectively, gas infalling on dynamical time scales ('cold mode'), adiabatic invariance 
during halo contraction, and detailed angular momentum conservation for the infalling gas.

The baryon fraction is assumed initially as the universal one, but the feedback--driven outflow 
reduces significantly this fraction. The parameters of the outflow model are fixed in such a way 
that namely the (low-mass) \ms--\mh\ relation at $z\sim 0$ is reproduced. 
In Firmani et al. (2010) we have shown that this happens only in the case of the 
so--called energy--driven outflow and for a high SN energy--transfer efficiency\footnote{If 
the possibility of late re--accretion of the lost gas is taken into account, 
then in Firmani et al. (2010) it was found that the momentum--driven outflow is the one 
that best can help to reproduce the $z=0$ \ms--\mh\ relation.}.
Because the exploration in this paper is only at the level of average trends, the halo MAHs used in 
the models are the averages among 20000 realizations for each given mass, and the used halo 
spin parameter $\lambda$ was fixed to 0.03 at all $z$'s, a value slightly smaller than the mean of 
$z=0$ relaxed halos in N-body cosmological simulations (e.g., Bett et al. 2007).
On the one hand, the N-body simulations show that $\lambda$ does not significantly changes with
time for halos growing in the accretion mode (e.g., Peirani, Mohayaee \& de Freitas Pacheco 2004;
D'Onghia \& Navarro 2007).
On the other hand, our models show that the radius evolution of disk galaxies agrees with 
observational inferences when the gas $\lambda$ parameter is roughly constant in time 
(Firmani \& Avila-Reese 2009). 

The GETs in Figure \ref{GHETs} show that the stellar mass growth is roughly proportional to 
the halo MAH as $z$ decreases. The model GETs and the GHETs in the \ms--\mh\ diagram (Fig. \ref{MsMh}) 
are very different; the average slopes of the GETs are $d$log\ms/$d$log\mh $\approx$1.4, much 
less than the slopes of the corresponding GHETs.
The model \ms-to-\mh\ ratio in Figure \ref{galefficiency} decreases toward the past due mainly to the 
feedback--driven outflow effect, but slightly, while the GHETs of the same masses evidence a very 
fast decreasing.
The \ms--\mh\ and (\ms/\mh)--\mh\ relations (Figs. \ref{MsMh} and \ref{galefficiency}, respectively) 
predicted by the model GETs at low masses change with $z$ (\ms\ slightly increases with $z$ for a 
fixed \mh) in an opposite direction with respect to the empirically inferred observations. 
This discrepancy represents a serious problem for all the theory behind the models.

The dramatic differences between the GETs and GHETs are also seen in the SSFR evolution plots 
(Figs. \ref{dotM} and \ref{SSFR-Ms}). We remark here the downsizing aspect. 
The GET-based SSFRs of the two galaxies evolve almost parallel (with a law
SSFR$\propto$ (1+$z$)$^{2.2-2.3}$), being the SSFR of the larger galaxy slightly higher since 
high $z'$s (weak upsizing). Instead, the GHET-based SSFRs cross each 
other at high $z$ (see the upper side of Figure \ref{dotM}), in such a way
that larger galaxies have higher SSFR at very high $z'$s, but at lower $z'$s,
the situation is inverted in such a way that as $z\rightarrow 0$, the smaller is the galaxy, 
the higher the SSFR (downsizing). The comparison of theoretical models with the GHET-based SSFRs 
suggests that some astrophysical mechanism not considered up to now should systematically 
delay the onset of SF activity the smaller the halo is.

The low SSFRs (early stellar mass assembly) of small galaxies predicted by the models, 
which include energy-driven outflows, are actually generic to all \lcdm-based models. 
This issue has been reported in different ways by the semi-analytic models. Most of these 
models show that the stellar population of small galaxies ($\ms\approx 10^9-10^{10.5}$ \msun) 
is assembled too early, becoming these galaxies older, redder, and with lower SSFRs at later 
epochs than the observed galaxies in the same mass range (e.g., Somerville et al. 2008; 
Fontanot et al. 2009; Santini et al. 2009). The problem is both at the 
level of satellite and central galaxies. For example, it was shown that models 
typically overpredict the observed stellar population ages (Pasquali et al. 2009) 
and stellar masses (Liu et al. 2010) of central galaxies in low mass halos.
By means of a galaxy evolutionary model similar to ours, Dutton, van den Bosch \& Dekel 
(2010b) also find that the SSFR of model low-mass galaxies is below the average
of observations, specially at redshifts $z\sim 1-2$, though these authors conclude
that their models are able to reproduce roughly the main features of the observed
SFR sequence.    

Finally, it should be said that N--body + hydrodynamics simulations of individual low mass
galaxies in a cosmological context also face the issue of too low SSFRs 
(as well as too high stellar and baryonic mass fractions; e.g., Col\'{\i}n et al. 
2010). In Firmani et al. (2010) we explored the inclusion in our semi-numeric models 
of re-accretion of the ejected gas for a broad range of possibilities. As expected, the 
SSFR of galaxies increases in general due to re-accretion but this increasing goes in 
the opposite direction of downsizing: only moderately for low-mass galaxies and too much 
for the larger galaxies. 

\section{Summary and Conclusions}

Empirically inferred \ms--\mh\ relations from $z\sim 0$ to $z\sim 4$ were connected with 
average \lcdm--halo MAHs in order to infer the corresponding individual {\it average} 
stellar mass growth histories, called here GHETs. 
We have adopted an \ms(\mh,$z$) relationship continuous in the $0-4$ redshift range and in 
agreement with the inferences by BCW10, who used the technique of jointly matching the 
abundances of observed \gsmfs\ to the theoretical $HMF$s at different $z'$s.
The main results obtained here, which allow to establish a unified description
of average galaxy evolutionary tracks for a large range of masses, are as follows.


$\bullet$ The folding in the \ms--\mh--$z$ surface, reflected mainly as peaks in 
the (\ms/\mh)--\mh\ diagram that shift to higher masses as $z$ increases, introduces 
an important feature in the behavior of the GHETs.
For masses much smaller than the peak mass at a given $z$, \Mmax($z$), 
the GHETs are in an active growth phase, whereas masses close to or greater than 
\Mmax($z$) are in a quiescent or completely passive (stagnated) 
phase (Figs. \ref{MsMh} and \ref{galefficiency}). Therefore, at each $z$ there is a 
characteristic stellar mass at which on average galaxies slow down their growth and 
transit from an active (star--forming blue) population to a passive (red) population. 
This transition mass decreases with time, giving rise to a phenomenon called here
{\it 'population downsizing'}.

$\bullet$ Galaxies less massive than \Mmax\ at $z=0$ (\ms$\lesssim 10^{10.5}$
\msun, \mh$\lesssim 10^{12.0}$ \msun) have still growing GHETs.
Besides, the lower the mass, the faster the later \ms\ growth, due likely to 
a delayed and lately active SF phase, a phenomenon called {\it 'downsizing in SSFR'}. 

$\bullet$ The shapes of the average stellar and halo mass assembling histories are quite different 
(Figure \ref{GHETs}). 
For galaxies that at $z=0$ have $\ms< 10^{10.5}$ \msun, their dark MAHs at later epochs 
grow slightly slower as smaller is the mass, while their stellar GHETs grow much faster. 
For $\ms(z=0)>10^{10.5}$ \msun, the larger the galaxy, the earlier its GHETs attain a stellar 
mass stagnation; until this epoch, the more massive the galaxy, the faster the \ms\ growth with 
respect to the corresponding \mh\ growth.

$\bullet$ By neglecting any stellar mass growth by accretion (dry mergers), 
a GHET allows to find the corresponding
SSFR history, SSFR($z$), by calculating \dotM\ and correcting by the stellar 
mas-loss recycling factor $R$. The inferred SSFRs at different $z$'s and as a function 
of \ms\ (Figs. \ref{dotM}, \ref{dotMobs}, and \ref{SSFR-Ms}) are reasonably consistent with 
recent observational studies based on direct measures of the SSFR of star-forming galaxies 
at different $z$'s. The GHET-based SSFRs corresponding to masses that at a given $z$ are
larger than the transition mass \mtran($z$), are much lower than the measured SSFRs of 
(rare) luminous blue galaxies, but agree or are even slightly larger than the measured
SSFRs of red galaxies. 
The overall consistency between the GHET-based SSFR-\ms\ relations at different $z'$s 
with direct inferences of these relations suggests that the accretion of stellar systems 
(mergers) plays a minor role in the \ms\ assembling of galaxies, excepting perhaps
those that transited to the passive sequence (the most massive ones).

$\bullet$ By using the SSFR vs. \ms\ evolutionary tracks, we calculated the characteristic 
transition mass at each $z$, \mtran($z$), above which the average GHET starts to significantly 
decrease its growth rate, transiting from the active to the passive galaxy population. 
The result is roughly described by the relation log(\mtran/\msun) = $10.30 + 0.55z$ (at least
up to $z\sim 2$). This result agrees with recent observational determinations of the evolution 
of the mass, at which the early- and late-type \gsmf\ components cross each other, 
from $z\approx 1$ to the present (Fig. \ref{Mtran}). 

$\bullet$ We determined also the transition rate in number density of active (blue) to quiescent (red) 
galaxy population. At $z\approx 0$ such a rate is $10^{-3.4\pm 0.2}$ gal Gyr$^{-1}$ Mpc$^{-3}$ and 
up to $z=1$ the rates are within $(1.0-5.5)\ 10^{-4}$ gal Gyr$^{-1}$ Mpc$^{-3}$, in good agreement with 
direct observational inferences of growth rate in number density of red galaxies. 

$\bullet$ We further explored whether \lcdm-based models of galaxy evolution are able to predict 
galaxy evolutionary tracks (called here 'GETs') in agreement with the GHETs, in particular at low masses 
where the downsizing in SSFR is evidenced by the GHETs.
We have shown that while the models with SN energy--driven outflows are able to reproduce the local 
\ms--\mh\ relation (see also Firmani et al. 2010), they fail in reproducing this relation at higher $z$'s. 
The difference between GETs and GHETs at low masses is rather large: the former ones show a fast 
decrease in SSFR with time almost independent of mass, while for the latter ones, the lower the mass, the
slower the late decrease of SSFR (downsizing in SSFR).


From these results, we conclude that the general description of stellar mass buildup of galaxies
provided by our average GHETs appears rather successful: the predicted SSFR histories as a function 
of \ms\ and the predicted transition mass \mtran\ as a function of $z$, as well as the transition
flux at different $z'$s, are roughly consistent with direct (but yet limited) observational 
estimates of all these quantities. 
Furthermore, our analysis reveals nicely the existence of a galaxy bi-modality related to the 
\ms\ growth activity: as $z$ decreases, smaller and smaller masses transit on average from the 
active (blue star-forming) to the passive (red) population. 

An important ingredient of our approach is the connection of \lcdm\ halos to the observed 
galaxies. Therefore, the underlying hierarchical \lcdm\ scenario is successful in the sense 
that the predictions of the model are in agreement with independent observations.
The average stellar and dark mass buildup of galaxies were found to be significantly 
different, specially for low and high masses. This result is in qualitative agreement with 
CW09 in spite of the differences in the data, method, and redshift range between their and our work.
However, the results from both works are different at a quantitative level, which is seen,
for example, in the different conclusions regarding the existence of a characteristic mass
at each epoch at which the SSFR of galaxies is truncated. 
Note that the SSFR and the determination of this characteristic mass (\mtran) are related to 
the first and second derivatives of the \ms\ growth, respectively. 

Our results put in a unified picture both the population downsizing (related to galaxies with 
$\ms\grtsim 3\ 10^{10}\msun$ at $z=0$) and the downsizing in SSFR (related to smaller galaxies). 
The challenges are now to explain the processes that produce: 
(1) the transition from active to passive regimes in \ms\ growth associated with a sharp cessation of the 
SFR in massive galaxies, 
(2) the fact that the typical mass of SSFR truncation, \mtran, decreases with cosmic time 
(population downsizing), and 
(3) the fact that the smaller is the halo, the more is delayed the galaxy's \ms\ growth
(downsizing in SSFR).

Items (1) and (2) seem to find partial explanations in aspects related to the same dark halo clustering 
(environment) evolution (e.g., Neistein et al. 2006) as well as to the introduction of feedback processes 
due to the AGNs of massive galaxies
(e.g., Silk \& Rees 1998; Kauffmann \& Haehnelt 2000; Granato et al. 2004; Cattaneo et al. 2005; 
de Lucia et al. 2005;  Bower et al. 2006; Croton et al. 2006).
Item (3), as shown here, is a sharp and less understood problem.
Later re-accretion of the gas ejected by galaxies due to SN-driven outflows does not account for a 
solution to this issue (Firmani et al. 2010). 
A better understanding of the astrophysical processes intervening in galaxy formation and evolution 
as well as the role of environmental processes is necessary. 

\acknowledgments

We are grateful to Dr. L. Pozzetti for kindly making available to us in electronic form the results 
reproduced in our Fig. \ref{Mtran}. We thank the anonymous referee for his/her careful review
of our paper. V. A. acknowledges PAPIIT-UNAM grant IN114509 and CONACyT grant 60354 for partial funding.

\end{document}